\documentclass[12pt,letterpaper]{article}

\usepackage{natbib,hyperref}
\usepackage[left=1in, top=1in, right=1in, bottom=1in]{geometry}
\usepackage{graphicx,bm,colonequals,amsmath,url}
\usepackage[font={footnotesize}]{caption,subcaption}
\usepackage{float}

\usepackage{animate,array,framed}
\newcounter{algorithm}
\newenvironment{algorithm}[1][]{\refstepcounter{algorithm}\par\medskip\noindent%
   \textbf{Algorithm~\thealgorithm: #1} \rmfamily}{\medskip}

\bibpunct[, ]{(}{)}{;}{a}{,}{,}

\newcommand{\bb}{\mathbf{b}}
\newcommand{\bs}{\mathbf{s}}

\newcommand{\bq}{\mathbf{q}}
\newcommand{\bx}{\mathbf{x}}
\newcommand{\by}{\mathbf{y}}
\newcommand{\bw}{\mathbf{w}}

\newcommand{\bz}{\mathbf{z}}

\newcommand{\bP}{\mathbf{P}}

\newcommand{\bJ}{\mathbf{J}}

\newcommand{\bI}{\mathbf{I}}

\newcommand{\bH}{\mathbf{H}}
\newcommand{\bU}{\mathbf{U}}
\newcommand{\bV}{\mathbf{V}}
\newcommand{\bK}{\mathbf{K}}

\newcommand{\bB}{\mathbf{B}}

\newcommand{\bR}{\mathbf{R}}

\newcommand{\bfzero}{\mathbf{0}}

\newcommand{\bfgamma}{\bm{\gamma}}

\newcommand{\bfxi}{\bm{\xi}}
\newcommand{\bftheta}{\bm{\theta}}
\newcommand{\bfeta}{\bm{\eta}}
\newcommand{\bfnu}{\bm{\nu}}
\newcommand{\bfdelta}{\bm{\delta}}

\newcommand{\bfbeta}{\bm{\beta}}

\newcommand{\bfSigma}{\bm{\Sigma}}
\newcommand{\bfOmega}{\bm{\Omega}}

\newcommand{\diag}{diag}
\newcommand{\tr}{tr}

\newcommand{\blockdiag}{blockdiag}

\newcommand{\knots}{\mathcal{W}}

\newcommand{\domain}{\mathcal{D}}

\newcommand{\dg}{\mbox{$^{\circ}$}}

\title{Parallel inference for massive distributed spatial data using low-rank models}

\author{Matthias Katzfuss\thanks{Department of Statistics, Texas A\&M University, \url{katzfuss@gmail.com}} \and Dorit Hammerling\thanks{Institute for Mathematics Applied to Geosciences, National Center for Atmospheric Research}}

\date{Published in Statistics and Computing (DOI: \href{http://dx.doi.org/10.1007/s11222-016-9627-4}{10.1007/s11222-016-9627-4})}

\begin{document}

\maketitle

\begin{abstract}
Due to rapid data growth, statistical analysis of massive datasets often has to be carried out in a distributed fashion, either because several datasets stored in separate physical locations are all relevant to a given problem, or simply to achieve faster (parallel) computation through a divide-and-conquer scheme. In both cases, the challenge is to obtain valid inference that does not require processing all data at a single central computing node.
We show that for a very widely used class of spatial low-rank models, which can be written as a linear combination of spatial basis functions plus a fine-scale-variation component, parallel spatial inference and prediction for massive distributed data can be carried out exactly, meaning that the results are the same as for a traditional, non-distributed analysis. The communication cost of our distributed algorithms does not depend on the number of data points. After extending our results to the spatio-temporal case, we illustrate our methodology by carrying out distributed spatio-temporal particle filtering inference on total precipitable water measured by three different satellite sensor systems.

\end{abstract}


\section{Introduction \label{sec:intro}}

While data storage capacity and data generation have increased by a factor of thousands in the past decade, the data transfer rate has increased by a factor of less than ten \citep{Zhang2013}.
It is therefore of increasing importance to develop analysis tools that minimize the movement of data and perform necessary computations in parallel where the data reside \citep[e.g.,][]{Fuller2011}. Here we consider two situations in which \emph{distributed data} can arise:
\begin{description}
\item[\textbf{Situation 1:}] Several massive datasets that are stored in separate data centers (servers) are all relevant to a given problem, and moving them to one central computing node for analysis is either too costly due to their large size or not desirable for other reasons such as unnecessary duplicated storage requirements. The goal then is to move the analysis to the data instead of the other way around \citep[e.g.,][]{Shoshani2010}.
\item[\textbf{Situation 2:}] All data relevant to a given problem are originally stored in the same location, but a ``divide-and-conquer'' approach with several nodes working in parallel on different chunks of the data is necessary, to achieve sufficiently fast computation or because the entire dataset is too large for a single machine to hold in working memory.
\end{description}
The goal in both of these situations is to obtain valid inference based on all data at a number of computers or servers, without moving the individual datasets between servers. The focus in this article is on Situation 1, but all results are also applicable to Situation 2 without modification.

In the spatial and environmental sciences, both of the described distributed-data situations arise frequently. Because analysis of a spatial dataset of size $n$ usually involves the data covariance matrix that has $n^2$ elements, Situation 2 applies to datasets of even moderate size. Situation 1 arises when several datasets containing information about a particular environmental variable are stored in different data centers throughout the US or the world, and we aim to obtain spatial inference and prediction based on all of them. For example, hundreds of millions of remotely sensed measurements of sea surface temperature per day are available both from the National Oceanic and Atmospheric Administration's (NOAA's) Advanced Very High Resolution Radiometer and from the National Aeronautics and Space Administration's (NASA's) Moderate Resolution Imaging Spectroradiometer. Measurements of column-integrated carbon dioxide are obtained by NASA's Orbiting Carbon Observatory-2 and Atmospheric InfraRed Sounder, Japan's Greenhouse Gases Observing Satellite, and other instruments.
With such satellite data, analyzing the data where they reside is especially important. It not only makes costly data movement and duplicate storage unnecessary, but also avoids (re)transfers of large amounts of data after changes in the retrieval algorithms, which occur quite regularly.
In this article, we will illustrate our methodology by making on-line spatio-temporal inference on a spatial variable called total precipitable water, based on measurements made by three major sensor systems stored at three associated data centers.

We consider here spatial low-rank models that consist of a component that can be written as a linear combination of spatial basis functions and a spatially independent fine-scale-variation term. Despite some recent criticism of their ability to approximate the likelihood of spatial processes with parametric covariances in certain low-noise situations \citep{Stein2013a}, low-rank models are a very widely used class of models for large spatial datasets (see Section \ref{sec:lowrank} below) because of their scalability for massive data sizes, and their predictive performance has been shown to compare favorably to other approaches in certain situations \citep{Bradley2014}. Note that here we do not advocate for or propose a particular spatial low-rank model --- rather, we are presenting distributed algorithms for inference that are applicable to all members of the class of spatial low-rank models.

We show that basic inference for low-rank models can be carried out \emph{exactly} for massive distributed spatial data, while only relying on (\emph{parallel}) \emph{local} computations at each server. In situations where a moderate, fixed number of basis functions is sufficient, the time complexity is linear in the number of measurements at each server, while the communication cost does not depend on the data size at all. Based on this main algorithm, we derive further algorithms for parameter inference and spatial prediction that are similarly well-suited for massive distributed data, and we extend the results to the spatio-temporal case. The results of our parallel distributed algorithms are exactly the same as those obtained by a traditional, non-distributed analysis with all data on one computational node, and so we do \emph{not} ignore spatial dependence between the data at different servers.

General-purpose computer-science algorithms for massive distributed data are not well suited to the distributed-spatial-data problem described above, as solving the linear systems required for prediction and likelihood evaluation would involve considerable movement of data or intermediary results. In the engineering literature, there has been some work on distributed Kalman filters for spatial prediction based on measurements obtained by robotic sensors \citep{Cortes2009, Xu2011, Graham2012}, but because the sensors are typically assumed to collect only one measurement at a time, we are not aware of any treatment of the case where the individual datasets are massive.

In the statistics literature, we are also not aware of previous treatment of the distributed-spatial-data problem of Situation 1, although it is possible to adapt some approaches proposed for analyzing (non-distributed) massive spatial data to the distributed case --- which is what we are doing with low-rank models in this article. The most obvious other approach is to simply approximate the likelihood for parameter estimation by dividing the data into blocks and then treating the blocks as independent, where in the distributed context each block would correspond to one of the distributed datasets. However, in most applications the distributed datasets were not necessarily collected in distinct spatial regions, and therefore block-independence approaches might ignore significant dependence between different blocks if there is substantial overlap in spatial coverage of the blocks. While methods such as composite likelihoods \citep[e.g.,][]{Vecchia1988,Curriero1999,Stein2004,Caragea2007,Caragea2008,Bevilacqua2012,Eidsvik2012} have been proposed to allow for some dependence between blocks, it is not clear how well these methods would work in our context, and how spatial predictions at unobserved locations should be obtained (e.g., to which block does the prediction location belong?). Other efforts to implement parallel algorithms for large spatial datasets \citep[e.g.,][]{Lemos2009} also exploit being able to split the data by spatial subregions and hence might not be suitable to distributed data in Situation 1.

This article is organized as follows. We begin with a brief review of low-rank spatial models in Section \ref{sec:lowrank}. We then focus on the distributed-data setting, describing a basic parallel algorithm for inference (Section \ref{sec:distributedinference}), discussing inference on model parameters and presenting important simplifications for fixed basis functions (Section \ref{sec:parameterinference}), describing how to do spatial prediction (Section \ref{sec:prediction}), and extending the methodology to the spatio-temporal setting (Section \ref{sec:spatiotemporal}). We present an application to total precipitable water measured by three sensor systems (Section \ref{sec:application}), and a simulation study exploring the effect of parallelization on computation time (Section \ref{sec:timing}). We conclude in Section \ref{sec:conclusions}.


\section{Spatial Low-Rank Models \label{sec:lowrank}}

\begin{figure}
\centering\includegraphics[width=.6\linewidth]{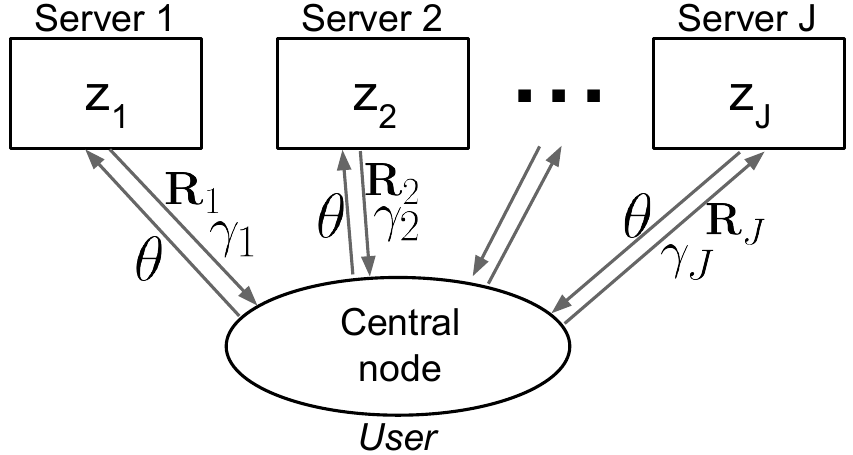}
\caption{An illustration of the set-up for distributed data with a central node and data stored at $J$ servers. The quantities to be transferred are described in Algorithm \ref{alg:spatial}.}
\label{fig:distributed_illustration}
\end{figure}

We are interested in making inference on a spatial process $\{y(\bs) \!: \bs \in \domain\}$, or $y(\cdot)$, on a continuous (non-gridded) domain $\domain$, based on a massive number of measurements, $\bz_{1:J} \colonequals (\bz_1',\ldots,\bz_J')'$, stored on $J$ different servers or data centers, where $\bz_j \colonequals (z(\bs_{j,1}),\ldots,z(\bs_{j,n_j}))'$ is stored on server $j$ (see Figure \ref{fig:distributed_illustration}), and the total number of measurements at locations $\{\bs_{j,i} \in \domain: i=1,\ldots,n_j; \, j=1,\ldots,J\}$ is given by $n \colonequals \sum_{j=1}^J n_j$. Note that the ordering of the servers is completely arbitrary and does not affect the results in any way. We assume that we have additive and spatially independent measurement error, such that
\begin{equation}
\label{eq:datamodel}
   z(\bs_{j,i}) = y(\bs_{j,i}) + \epsilon(\bs_{j,i}),
\end{equation}
for all $i=1,\ldots,n_j$ and $j=1,\ldots,J$, where $\epsilon(\bs_{j,i}) \sim N(0, v_\epsilon(\bs_{j,i}))$ is independent of $y(\cdot)$, and the function $v_\epsilon(\cdot)$ is known. 
In practice, if $v_\epsilon(\cdot)$ is unknown, one can set $v_\epsilon(\cdot) \equiv \sigma^2_\epsilon$, and then estimate $\sigma^2_\epsilon$ by extrapolating the variogram to the origin \citep{Kang2009}. Because the measurements in \eqref{eq:datamodel} are at point level and not on a grid, we assume for simplicity that no two measurement locations coincide exactly.

The true process $y(\cdot)$ is assumed to follow a spatial low-rank model of the form,
\begin{equation}
\label{eq:lowrankmodel}
  y(\bs) = \bb(\bs)'\bfeta + \delta(\bs), \quad \bs \in \domain,
\end{equation}
where $\bb(\cdot)$ is a vector of $r$ spatial basis functions with $r<<n$, $\bfeta \sim N_r(\bfnu_0,\bK_0)$, and often $\bfnu_0 = \bfzero$. The fine-scale variation $\delta(\bs) \sim N(0,v_\delta(\bs)  )$ is spatially independent and independent of $\bfeta$. Note that we did not include a spatial trend in \eqref{eq:lowrankmodel}, as any linear trend of the form $\bx(\cdot)'\bfbeta$, where $\bx(\cdot)$ is a vector of spatial covariates, can simply be absorbed into $\bb(\cdot)'\bfeta$ if we assign a normal prior distribution to $\bfbeta$.

Low-rank models of the form \eqref{eq:lowrankmodel} are popular because they do not assume stationarity, and, for a fixed number of basis functions, the time complexity to obtain exact spatial predictions is linear in the number of measurements, hence offering excellent scalability for massive datasets. Many widely used classes of spatial models are of the form \eqref{eq:lowrankmodel}, such as the spatial random effects model \citep{Cressie2008}, discretized convolution models \citep[e.g.,][]{Higdon1998,Calder2007,Lemos2009}, and the predictive process \citep{Banerjee2008,Finley2009}. Basis functions that have been used in \eqref{eq:lowrankmodel} include empirical orthogonal functions \citep[e.g.][]{Mardia1998,Wikle1999}, Fourier basis functions \citep[e.g.,][]{Xu2005}, W-wavelets \citep[e.g.,][]{Shi2007,Cressie2010a,Kang2009a}, and bisquare functions \citep[e.g.,][]{Cressie2008,Katzfuss2010,Katzfuss2011}.

Note that our methodology described in the following sections is applicable to any of these parameterizations of \eqref{eq:lowrankmodel}, and we do not advocate for a particular model over others. Hence, we work with the general class of spatial low-rank models in \eqref{eq:lowrankmodel}, and we only assume that there is some parameter vector, $\bftheta$, that determines $\bb(\cdot)$, $\bK_0$, and $v_\delta(\cdot)$.


\section{Distributed Spatial Inference --- Main Algorithm \label{sec:distributedinference}}

We will now discuss how to obtain the posterior distribution, $[\bfeta | \bz_{1:J}]$, of the random-effects vector of basis-function weights, $\bfeta$, by performing parallel computations at each server $j$ that use only the local data $\bz_j$. Throughout this section, we will treat the parameter vector $\bftheta$ as fixed and known, with parameter inference to be discussed in Section \ref{sec:parameterinference}.

First, define $\bB_{1:J} = (\bB_1',\ldots,\bB_J')'$ and $\bV_{1:J}=\blockdiag(\bV_1,\ldots,\bV_J)$, where $\bz_j$, $\bB_j \colonequals (\bb(\bs_{j,1}),\ldots,\bb(\bs_{j,n_j}))'$, and $\bV_j \colonequals \diag(v_\delta(\bs_{j,1})\!+\!v_\epsilon(\bs_{j,1}),\ldots,v_\delta(\bs_{j,n_j})\!+\!v_\epsilon(\bs_{j,n_j}))$ are the local quantities at server $j$. Viewing the random-effects vector $\bfeta$ as a Bayesian parameter with prior $\bfeta \sim N_r(\bfnu_0,\bK_0)$ and linear Gaussian ``likelihood'' $\bz_{1:J} | \bfeta \sim N_n(\bB_{1:J}\bfeta,\bV_{1:J})$, it is easy to see that the posterior distribution of $\bfeta$ given the data at all servers is also multivariate normal, $\bfeta | \bz_{1:J} \sim N_r(\bfnu_{z}, \bK_{z} )$, where 
\begin{equation*}
\label{eq:posteriors}
\begin{split}
  \bK_z^{-1} &= \bK_{0}^{-1} + \bR, \quad \bR \colonequals  \bB_{1:J}'\bV_{1:J}^{-1} \bB_{1:J}\\
  \bfnu_z & = \bK_z (\bK_0^{-1} \bfnu_0 + \bfgamma), \quad \bfgamma \colonequals \bB_{1:J}'\bV_{1:J}^{-1} \bz_{1:J}.
\end{split}
\end{equation*}
The key to our distributed algorithms is that, due to the diagonal block structure of $\bV_{1:J}$, we have
\begin{equation}
\label{eq:infofilter}
\begin{split}
 \bR & = \textstyle\sum_{j=1}^J\bB_j'\bV_j^{-1} \bB_j  \equalscolon \textstyle\sum_{j=1}^J \bR_j\\
 \bfgamma & = \textstyle\sum_{j=1}^J\bB_j'\bV_j^{-1} \bz_j  \equalscolon \textstyle\sum_{j=1}^J \bfgamma_j.
\end{split}
\end{equation}
Thus, the posterior distribution of $\bfeta$ can be obtained by properly combining quantities that each only depend on the data and their spatial locations at one of the servers. This implies the following parallel algorithm to obtain the posterior distribution of $\bfeta$:
\begin{framed}
\begin{algorithm}
\label{alg:spatial} \textbf{Distributed Spatial Inference}
\begin{enumerate}
\item Do the following \emph{in parallel} for $j=1,\ldots,J$:
 \begin{enumerate}
  \item Move $\bftheta$ to server $j$ (where data $\bz_j$ is stored) and create the matrices $\bB_j$ and $\bV_j$ there.
  \item At server $j$, calculate $\bR_j = \bB_j'\bV_j^{-1} \bB_j$ and $\bfgamma_j = \bB_j'\bV_j^{-1}\bz_j$.
  \item Transfer the $r \times r$ matrix $\bR_j$ and the $r\times 1$ vector $\bfgamma_j$ back to the central node.
 \end{enumerate}
\item At the central node, calculate $\bK_z^{-1} = \bK_{0}^{-1} + \textstyle\sum_{j=1}^J \bR_j$ and $\bfnu_z = \bK_z (\bK_0^{-1} \bfnu_0 + \textstyle\sum_{j=1}^J \bfgamma_j)$. The posterior distribution of $\bfeta$ is given by $\bfeta | \bz_{1:J} \sim N_r(\bfnu_{z}, \bK_{z} )$.
\end{enumerate}
\end{algorithm}
\end{framed}
Algorithm \ref{alg:spatial} is illustrated in Figure \ref{fig:distributed_illustration}. The overall time complexity is $\mathcal{O}(r^3 +  r^2 max_j n_j)$ (specifically, $\mathcal{O}(r^3)$ at the central node and $\mathcal{O}(r^2n_j)$ at server $j$), the memory complexity is $\mathcal{O}(r^2)$ at the central node and $\mathcal{O}(r n_j)$ at server $j$, and we need to move only the $r(r/2 +3/2)$ unique elements in $\bR_j$ and $\bfgamma_j$ from each server. Compare this to a non-distributed algorithm that has time complexity $\mathcal{O}(r^3 + r^2\sum_j n_j)$, memory complexity $\mathcal{O}(rn)$, and requires moving the $n$ measurements (plus their spatial coordinates) to the central node. In summary, if $r$ remains fixed as the data size increases, Algorithm \ref{alg:spatial} has computational cost that is linear in each $n_j$, the communication cost does not depend on $n$ at all, and hence it is scalable for massive distributed datasets.

\subsection{Reducing Communication Via Sparsity \label{sec:sparsity}}

The required amount of communication and computation for Algorithm \ref{alg:spatial} can be reduced further if the basis-function matrices $\bB_j$ are sparse, resulting in sparse $\bR_j$. Several approaches that impose sparsity in basis-function models have recently been proposed \citep[e.g.,][]{Lindgren2011a,Nychka2012}. While sparsity in principle allows fast computation even for large $r$, having $r = \mathcal{O}(n)$ does not allow the reduction in communication desired in our Situation 1 from Section \ref{sec:intro}. Large $r$ also creates problems in the spatio-temporal filtering context described below in Section \ref{sec:spatiotemporal}, as sparsity in the precision matrix $\bK_z^{-1}$ can generally not be maintained after propagation through time.

Returning to the low-rank case with small to moderate $r$, sparsity can be achieved, for example, by taking the predictive-process approach \citep[see ][for a detailed definition]{Banerjee2008} with a compactly supported parent covariance function, as follows. Assume a set of knots, $\knots \colonequals \{ \bw_1, \ldots, \bw_r \}$, and a parent covariance function
\[
C(\bs_1,\bs_2) = \sigma(\bs_1) \sigma(\bs_2) \rho(\bs_1,\bs_2), \quad \bs_1,\bs_2 \in \domain,
\]
where $\rho$ is a correlation function. Then the predictive process can be written in the form \eqref{eq:lowrankmodel} with
\begin{equation}
\label{eq:ppspatial}
 \bb(\bs) \colonequals \sigma(\bs) \,\big(\rho( \bs, \bw_1),\ldots, \rho( \bs, \bw_r)\big)', \; \bs \in \domain,
\end{equation}
and the $(i,j)$th element of $\bK_0^{-1}$ given by $\rho( \bw_i, \bw_j)$ \citep[see, e.g.,][]{Katzfuss2012}.

Now, if $C$ is compactly supported with range $h$, then the $(l,m)$th element of the matrix $\bR_j = \bB_j'\bV_j^{-1}\bB_j$ in \eqref{eq:infofilter} can only be nonzero if $||\bw_l-\bw_m||<2h$. Hence, if for a given set of knots, at most $v$ other knots are within a distance of $2h$ of any knot, at most $r( v/2 +2)$ numbers (including $\bfgamma_j$) need to be transferred from each server.


\section{Parameter Inference \label{sec:parameterinference}}

So far, we have treated the parameter vector $\bftheta$ (containing the parameters determining $\bb(\cdot)$, $\bK_0$, and $v_\delta(\cdot)$) as fixed and known. In practice, of course, this is usually not the case. Fortunately, several commonly used inference approaches can be implemented in a distributed and parallel fashion by extending Algorithm \ref{alg:spatial} (while still producing the same results as in the traditional, non-distributed setting).

\subsection{Parsimonious Parameterizations \label{sec:parsparam}}

If the parameter vector $\bftheta$ is of low dimension (e.g., there are only three parameters in the predictive-process model in \eqref{eq:ppspatial} with a Mat\'ern parent covariance function), and estimates or posterior distributions of the parameters are not available in closed form, standard numerical likelihood-based inference is one possibility for parameter inference.

As shown in Appendix \ref{appendix}, the likelihood (up to a normalization constant) for the spatial low-rank model in Section \ref{sec:lowrank} can be written as,
\begin{equation}
\label{eq:distributedlikelihood}
 -2 \log L(\bftheta) \colonequals -2 \log [\bz_{1:J} | \bftheta ] =   - \log |\bK_0^{-1}| + \bfnu_0'\bK_0^{-1}\bfnu_0 + \log |\bK_z^{-1}| - \bfnu_z'\bK_z^{-1}\bfnu_z + \textstyle\sum_{j=1}^J a_j,
\end{equation}
where $a_j \colonequals \log|\bV_j| + \bz_j'\bV_j^{-1} \bz_j$. This allows carrying out both frequentist and Bayesian inference for distributed data (e.g., by numerical maximization of the likelihood, Metropolis-Hasting sampling, or other approaches). Each iteration of such a parameter-inference procedure consists of carrying out Algorithm \ref{alg:spatial} (with the addition of calculating $a_j$ at server $j$ and moving this scalar quantity to the central node), combining the results to evaluate the likelihood \eqref{eq:distributedlikelihood} at the central node, updating the parameters $\bftheta$, and sending out the new value of $\bftheta$ to the servers. This results in a sequential algorithm, for which the (major) calculations at each iteration can be carried out in parallel.

To avoid servers being idle in such a sequential algorithm, we recommend instead the use of an importance or particle sampler. Any of the various such algorithms proposed in the literature can be carried out in the distributed context (with the exact same results), by evaluating the likelihood as in \eqref{eq:distributedlikelihood}. Here is an example of such an algorithm:
\begin{framed}
\begin{algorithm}
\label{alg:impsampler} \textbf{Distributed Importance Sampler}
\begin{enumerate}
\item Generate a number of parameter vectors or particles, $\bftheta^{(1)},\ldots,\bftheta^{(M)}$, from a suitably chosen proposal distribution, $q(\bftheta)$.
\item Do the following \emph{in parallel} for $j=1,\ldots,J$ and $m=1,\ldots,M$:
 \begin{enumerate}
  \item Move $\bftheta^{(m)}$ to server $j$ and create the matrices $\bB_j^{(m)}$ and $\bV_j^{(m)}$.
  \item Calculate 
\begin{align*}
\bR_j^{(m)} & = \bB_j^{(m)}{}'(\bV_j^{(m)})^{-1} \bB_j^{(m)}\\
\bfgamma_j^{(m)} & = \bB_j^{(m)}{}'(\bV_j^{(m)})^{-1}\bz_j\\
a_j^{(m)} & =\log|\bV_j^{(m)}| + \bz_j'(\bV_j^{(m)})^{-1} \bz_j.
\end{align*}
  \item Transfer $\bR_j^{(m)}$, $\bfgamma_j^{(m)}$, and $a_j^{(m)}$ back to the central node.
 \end{enumerate}
\item At the central node, for $m=1,\ldots,M$, calculate $(\bK_z^{(m)})^{-1} =(\bK_{0}^{(m)})^{-1} + \textstyle\sum_{j=1}^J \bR_j^{(m)}$, $\bfnu_z^{(m)} = \bK_z^{(m)} ((\bK_0^{(m)})^{-1} \bfnu_0^{(m)} + \textstyle\sum_{j=1}^J \bfgamma_j^{(m)})$, and 
\begin{align*}
 -2 \log L(\bftheta^{(m)})  & = - \log |(\bK_0^{(m)})^{-1}| + \bfnu_0^{(m)}{}'(\bK_0^{(m)})^{-1}\bfnu_0^{(m)} \\
& \qquad + \log |(\bK_z^{(m)})^{-1}| - \bfnu_z^{(m)}{}'(\bK_z^{(m)})^{-1}\bfnu_z^{(m)} + \textstyle\sum_{j=1}^J a_j^{(m)}.
\end{align*}
\item The particle approximation of the posterior distribution of $\bftheta$ takes on the value $\bftheta^{(m)}$ with probability $w^{(m)} \propto p(\bftheta^{(m)}) L(\bftheta^{(m)})/q(\bftheta^{(m)})$ for $m=1,\ldots,M$, where $p(\bftheta)$ is the prior distribution of the parameters.
\end{enumerate}
\end{algorithm}
\end{framed}
The advantage of this parameter-inference approach is that we can carry out calculations for the likelihood evaluations for all particles completely in parallel at all servers (while getting the same results as in the traditional, non-distributed setting).

\subsection{Spatial Random Effects Model \label{sec:sreinference}}

The spatial random effects model \citep{Cressie2008,Katzfuss2009,Kang2009a} is a low-rank model of the form \eqref{eq:datamodel}--\eqref{eq:lowrankmodel}, for which the basis functions are known functions (e.g., bisquare functions) that do not depend on unknown parameters, $\bK_0$ is a general covariance matrix (i.e., it contains $r(r+1)/2$ parameters), and often $v_\delta(\cdot) \equiv \sigma^2_\delta$. If we also assume $v_\epsilon(\cdot) \equiv \sigma^2_\epsilon$ (or we have transformed the data such that these assumptions hold), we have $\bV_j^{-1} = \frac{1}{\sigma^2_\delta + \sigma^2_\epsilon} \bI_{n_j}$, and so $\bR_j = \frac{1}{\sigma^2_\delta + \sigma^2_\epsilon}\bB_j'\bB_j$ and $\bfgamma_j = \frac{1}{\sigma^2_\delta + \sigma^2_\epsilon} \bB_j'\bz_j$. Since the $\bB_j$ in the spatial random effects model are fixed, all that is required for inference on $\bfeta$ in Algorithm \ref{alg:spatial} from server $j$ are the fixed quantities $\bB_j'\bz_j$ and $\bB_j'\bB_j$, making multiple passes over the servers for parameter inference unnecessary. The only additional information required from server $j$ for evaluating the likelihood \eqref{eq:distributedlikelihood} is $n_j$ and $\bz_j'\bz_j$.

If the basis functions do contain unknown parameters, or $v_\epsilon(\cdot)$ is not constant, maximum likelihood estimates can be obtained by deriving a distributed version of the expectation-maximization algorithm of \citet{Katzfuss2009,Katzfuss2010}. Each step of the resulting algorithm consists of carrying out Algorithm \ref{alg:spatial}, and then updating the estimates of $\bK_0^{-1}$ and $\sigma_\delta^2$ as
\begin{align*}
\label{eq:emupdate}
\hat\bK_0^{-1} & = (\bK_z + \bfnu_z \bfnu_z')^{-1} = \bK_z^{-1} - \bq\bq'/(1+\bfnu_z'\bq)\\
\hat\sigma^2_\delta & = \sigma^2_\delta + \sum_{j=1}^J \frac{\sigma^4_\delta}{n_j} \big( || \bV_j^{-1}(\bz_j - \bB_j \bfnu_z)||^2 - \tr(\bfOmega_j^{-1}) \big),
\end{align*}
where $\bq \colonequals \bK_0^{-1} \bfnu_0 + \bfgamma$ and $\bfOmega_j \colonequals \bB_j \bK_{z} \bB_j' + \bV_j$. The expression for $\hat\sigma^2_\delta$  above can be derived by obtaining $[\bfdelta_j | \bfeta, \bz_{1:J}]$ and then applying the laws of total expectation and total variance.

By assuming conjugate prior distributions (i.e., an inverse-Wishart distribution for $\bK_0$ and an inverse-Gamma distribution for $\sigma^2_\delta$), Bayesian inference using a Gibbs sampler is also possible.


\section{Spatial Prediction \label{sec:prediction}}

The goal in spatial statistics is typically to make spatial predictions of $y(\cdot)$ at a set of prediction locations, $\bs_1^P,\ldots,\bs^P_{n_P}$, based on all data $\bz_{1:J}$, which in technical terms amounts to finding the posterior predictive distribution $[\by^P | \bz_{1:J}]$, where $\by^P \colonequals (y(\bs_1^P),\ldots,y(\bs^P_{n_P}))'$. Note that prediction can be carried out separately, after parameter inference has been completed, and so it suffices to obtain the predictive distribution for the final parameter estimates in a frequentist procedure, or for thinned MCMC samples or for particles with nonzero weight in a Bayesian context. 

Because we can write
\begin{equation}
\label{eq:spatialpred}
  \by^P = \bB^P \bfeta + \bfdelta^P,
\end{equation}
where $\bB^P \colonequals (\bb(\bs^P_{1}),\ldots,\bb(\bs^P_{n_P}))'$ and $\bfdelta^P \colonequals (\delta(\bs_1^P),\ldots,\delta(\bs^P_{n_P}))'$, the desired predictive distribution is determined by the joint posterior distribution $[\bfeta, \bfdelta^P | \bz_{1:J}]$. 

First, assume that none of the prediction locations exactly coincide with any of the observed locations. This is a reasonable assumption when measurements have point support on a continuous spatial domain, as we have assumed throughout this manuscript. Then it is easy to see that $\bfdelta^P | \bz_{1:J} \sim N_{n_P}(\bfzero,\bV_\delta^P)$, with $\bV_\delta^P \colonequals \diag\{v_\delta(\bs^P_{1}),\ldots,v_\delta(\bs^P_{n_P})\}$, is conditionally independent of $\bfeta$ given $\bz_{1:J}$. Therefore, spatial prediction reduces to obtaining $\bfnu_z$ and $\bK_z$ using Algorithm \ref{alg:spatial}, and then calculating 
\begin{equation}
\label{eq:prediction2}
  \by^P | \bz_{1:J} \sim N_{n_P}( \bB^P \bfnu_z, \bB^P \bK_z \bB^P{}' + \bV_\delta^P )
\end{equation}
at the central node.

Appendix \ref{sec:predoverlap} describes how to do spatial prediction when a small number of the observed locations coincide with the desired prediction locations.


\section{Spatio-Temporal Inference \label{sec:spatiotemporal}}

To extend our results to the spatio-temporal case, we consider a spatio-temporal low-rank model in discrete time. In our hierarchical state-space model, the process of interest is given by,
\[
 y_t(\bs) = \bb_t(\bs)'\bfeta_t + \delta_t(\bs), \quad \bs \in \domain; \; t=1,2,\ldots,
\]
where $\delta_t(\cdot)$ is assumed to be independent over space and time with variance function $v_{\delta,t}(\cdot)$, and the temporal evolution of the low-rank component is given by,
\[
  \bfeta_t | \bftheta_t, \bfeta_{t-1}, \bfeta_{t-2},\ldots \sim N_r(\bH_t \bfeta_{t-1}, \bU_t), \quad t=1,2,\ldots,
\]
where $\bfeta_0 \sim N_r(\bfnu_{0,0},\bK_{0,0})$ is the initial state, and $\bftheta_t$ is a time-varying parameter vector with generic transition equation $p(\bftheta_t|\bftheta_{t-1})$ and initial value $\bftheta_0$. The data at server $j$ at time $t$ are given by $\bz_{j,t} \colonequals (z_t(\bs_{1,j,t}),\ldots,z_t(\bs_{n_{j,t},j,t}))'$, with
\[
  z_{t}(\bs_{i,j,t}) = y_{t}(\bs_{i,j,t}) + \epsilon_{t}(\bs_{i,j,t}),
\]
for all $i=1,\ldots,n_{j,t}$, $j=1,\ldots,J$, and $t=1,2,\ldots$, where $\epsilon_{t}(\bs_{i,j,t}) \sim N(0, v_\epsilon(\bs_{i,j,t}))$ is independent in space, time, and of $y(\cdot)$. 
\citet{Cressie2010a} called this the spatio-temporal random effects model, but as in the spatial-only case, many different ways of parameterizing such a spatio-temporal low-rank model are possible (see Section \ref{sec:application} for an example). We again merely assume that $\bH_t$, $\bU_t$, $\bb_t(\cdot)$, and $v_{\delta,t}(\cdot)$ are known up to the parameter vector $\bftheta_t$.

\subsection{Filtering and Smoothing for Known Parameters}

We temporarily assume the parameters $\bftheta_1,\bftheta_2,\ldots$ to be known, or held at a particular set of values at one step of a parameter-inference procedure (see Section \ref{sec:STparam} below). We first take an on-line, filtering perspective in time, which means that we are interested at time point $t$ in obtaining the filtering distribution $\bfeta_t | \bz_{1:t} \sim N_r(\bfnu_{t|t},\bK_{t|t})$, where $\bz_{1:t}$ denotes the vector of all data collected at the first $t$ time points. We can obtain $\bfnu_{t|t}$ and $\bK_{t|t}$ using a Kalman filter, for which each update step essentially requires carrying out Algorithm \ref{alg:spatial}:
\begin{framed}
\begin{algorithm}
\label{alg:filtering} \textbf{Distributed Spatio-Temporal Filtering}
\begin{enumerate}
\item For $t=0$, initialize the algorithm by calculating $\bfnu_{0|0}$ and $\bK_{0|0}$ based on $\bftheta_0$.
\item At time $t=1,2,\ldots$, once the new data $\bz_{1,t},\ldots,\bz_{J,t}$ become available:
 \begin{enumerate}
\item Do the following \emph{in parallel} for $j=1,\ldots,J$:
 \begin{enumerate}
  \item Move $\bftheta_t$ to server $j$ and create the matrices $\bB_{j,t}$ and $\bV_{j,t}$ based on the observed locations at time $t$.
  \item At server $j$, calculate $\bR_{j,t} = \bB_{j,t}'\bV_{j,t}^{-1} \bB_{j,t}$ and $\bfgamma_{j,t} = \bB_{j,t}'\bV_{j,t}^{-1}\bz_{j,t}$.
  \item Transfer $\bR_{j,t}$ and $\bfgamma_{j,t}$ back to the central node.
 \end{enumerate}
\item At the central node, calculate the forecast quantities $\bfnu_{t|t-1} \colonequals \bH_t \bfnu_{t-1|t-1}$, $\bK_{t|t-1} \colonequals \bH_t \bK_{t-1|t-1} \bH_t' + \bU_t$, and then the filtering quantities $\bK_{t|t}^{-1} = \bK_{t|t-1}^{-1} + \textstyle\sum_{j=1}^J \bR_{j,t}$ and $\bfnu_{t|t} = \bK_{t|t} (\bK_{t|t-1}^{-1} \bfnu_{t|t-1} + \textstyle\sum_{j=1}^J \bfgamma_{j,t})$. We have $\bfeta_t | \bz_{1:t} \sim N_r(\bfnu_{t|t},\bK_{t|t})$.
\end{enumerate}
\end{enumerate}
\end{algorithm}
\end{framed}
It is interesting to note that Algorithm \ref{alg:spatial} in Section \ref{sec:distributedinference} can itself be viewed as a decentralized Kalman filter \citep{Rao1993} over servers applied to our spatial low-rank model written as a state-space model with an identity evolution equation. Thus, Algorithm \ref{alg:filtering} is actually the combination of two nested filters, where each ``outer'' filtering step over time essentially consists of an ``inner'' filter over servers as in \eqref{eq:infofilter}.

In some applications, retrospective smoothing inference based on data collected at $T$ time points might be of interest. Obtaining the smoothing distribution $\bfeta_t | \bz_{1:T} \sim N_r(\bfnu_{t|T},\bK_{t|T})$ for $t=1,\ldots,T$, requires forward-filtering using Algorithm \ref{alg:filtering} and then backward-smoothing at the central node by calculating iteratively for $t=T-1,T-2,\ldots,1$:
\begin{align*}
\bfnu_{t|T} & = \bfnu_{t|t} + \bJ_t(\bfnu_{t+1|T} - \bfnu_{t+1|t}),\\
\bK_{t|T} & = \bK_{t|t} + \bJ_t(\bK_{t+1|T} - \bK_{t+1|t})\bJ_t',
\end{align*}
where $\bJ_t \colonequals \bK_{t|t}\bH_{t+1}'\bK_{t+1|t}^{-1}$ \citep[see, e.g.,][p.~732, for more details]{Cressie2010a}. Also, note that in the smoothing context, it is not actually necessary to ``consolidate'' the information at the end of each time point as in Step 2(b) of Algorithm \ref{alg:filtering} before moving on to the next time point; instead, we can calculate $\bR_{j,1},\ldots,\bR_{j,T}$ and $\bfgamma_{j,1},\ldots,\bfgamma_{j,T}$ at each server $j$, and then directly calculate $\bK_{T|T}$ and $\bfnu_{T|T}$ at the central node.

Because $\delta_t(\cdot)$ is \emph{a priori} independent over time, spatial prediction for each $t$ in the filtering and smoothing context can be carried out as described in Section \ref{sec:prediction} using the filtering or smoothing distribution of $\bfeta_t$ (i.e., $\bfnu_{t|t}, \bK_{t|t}$ or $\bfnu_{t|T}, \bK_{t|T}$, respectively).

\subsection{Spatio-Temporal Parameter Inference \label{sec:STparam}}

In the filtering context, inference on the parameter vector $\bftheta_t$ at time point $t$ is typically based on the filtering likelihood,
\begin{equation}
\label{eq:filtlikelihood}
\begin{split}
-2 \log L_t(\bftheta_t) & \colonequals -2 \log [\bz_t | \bz_{1:t-1}, \bftheta_t ]   \\
& = - \log |\bK_{t|t-1}^{-1}| + \bfnu_{t|t-1}'\bK_{t|t-1}^{-1}\bfnu_{t|t-1} + \log |\bK_{t|t}^{-1}| - \bfnu_{t|t}'\bK_{t|t}^{-1}\bfnu_{t|t} + \textstyle\sum_{j=1}^J a_{j,t},
\end{split}
\end{equation}
where $a_{j,t} \colonequals  \log|\bV_{j,t}| + \bz_{j,t}'\bV_{j,t}^{-1} \bz_{j,t}$. This expression of the likelihood can be derived similarly as in the spatial-only case described in Appendix \ref{appendix}.
If there are a small number of unknown parameters in the spatio-temporal low-rank model, we again advocate the use of a particle-filtering approach for parameter estimation. Sequential importance sampling with resampling \citep{Gordon1993} is a natural inference procedure for on-line inference over time. With distributed data, it can be carried out using a straightforward combination of Algorithms \ref{alg:impsampler} and \ref{alg:filtering}:
\begin{framed}
\begin{algorithm}
\label{alg:particlefilter} \textbf{Distributed Spatio-Temporal Particle Filter}
\begin{enumerate}
\item For $t=0$, calculate $\bfnu_{0|0}$ and $\bK_{0|0}$ based on initial parameter value $\bftheta_0$. Then sample $M$ particles $\bftheta_1^{(1)},\ldots,\bftheta_1^{(M)}$ from a suitably chosen proposal distribution $q(\bftheta_1|\bftheta_0)$.
\item At time $t=1,2,\ldots$, once new data $\bz_{1,t},\ldots,\bz_{J,t}$ become available:
 \begin{enumerate}
\item Do the following \emph{in parallel} for $j=1,\ldots,J$ and $m=1,\ldots,M$:
 \begin{enumerate}
  \item Move $\bftheta_t^{(m)}$ to server $j$ and create the matrices $\bB_{j,t}^{(m)}$ and $\bV_{j,t}^{(m)}$ based on the observed locations at time $t$.
  \item At server $j$, calculate 
\begin{align*}
\bR_{j,t}^{(m)} &= \bB_{j,t}^{(m)}{}'(\bV_{j,t}^{(m)})^{-1} \bB_{j,t}^{(m)}\\
\bfgamma_{j,t} & = \bB_{j,t}^{(m)}{}'(\bV_{j,t}^{(m)})^{-1}\bz_{j,t}\\
a_{j,t}^{(m)} & =\log|\bV_{j,t}^{(m)}| + \bz_{j,t}'(\bV_{j,t}^{(m)})^{-1} \bz_{j,t}.
\end{align*}
  \item Transfer $\bR_{j,t}^{(m)}$, $\bfgamma_{j,t}^{(m)}$, $a_{j,t}^{(m)}$ and back to the central node.
 \end{enumerate}
\item At the central node, do the following in parallel for $m=1,\ldots,M$:
 \begin{enumerate}
  \item Based on $\bftheta^{(m)}$, calculate $\bfnu_{t|t-1}^{(m)} \colonequals \bH_t^{(m)} \bfnu_{t-1|t-1}^{(m)}$, \\
$\bK_{t|t-1}^{(m)} \colonequals \bH_t^{(m)} \bK_{t-1|t-1}^{(m)} \bH_t^{(m)}{}' + \bU_t^{(m)}$, 
$(\bK_{t|t}^{(m)})^{-1} = (\bK_{t|t-1}^{(m)})^{-1} + \textstyle\sum_{j=1}^J \bR_{j,t}^{(m)}$, 
$\bfnu_{t|t}^{(m)} = \bK_{t|t}^{(m)} ((\bK_{t|t-1}^{(m)})^{-1} \bfnu_{t|t-1}^{(m)} + \textstyle\sum_{j=1}^J \bfgamma_{j,t}^{(m)})$, and the filtering likelihood $L_t(\bftheta_t^{(m)})$ as in \eqref{eq:filtlikelihood}.
 \end{enumerate}
\item The particle-filter approximation of the filtering distribution of $\bftheta_t$ takes on the value $\bftheta_t^{(m)}$ with probability $w_t^{(m)} \propto p(\bftheta_t^{(m)}|\bftheta_{t-1}^{(m)}) L_t(\bftheta_t^{(m)})/q(\bftheta_t^{(m)}|\bftheta_{t-1}^{(m)})$.
\item Using a resampling scheme \citep[see, e.g.,][]{Douc2005}, generate resampled particles $\widetilde{\bftheta}_t^{(1)},\ldots,\widetilde{\bftheta}_t^{(M)}$ (and the associated $\bK_{t|t}^{(m)}$ and $\bfnu_{t|t}^{(m)}$) from $\bftheta_t^{(1)},\ldots,\bftheta_t^{(M)}$, and obtain $M$ particles for time $t+1$ using a suitable proposal distribution $q(\bftheta_{t+1}|\widetilde{\bftheta}_t^{(m)})$.
\end{enumerate}
\end{enumerate}
\end{algorithm}
\end{framed}


In a smoothing context, parameter inference is based on the likelihood, $[\bz_{1:T} | \bftheta_{1:T} ] = \prod_{t=1}^T [\bz_t | \bz_{1:t-1}, \bftheta_t ]$, of all data in a specific time window $\{1,\ldots,T\}$, where $[\bz_t | \bz_{1:t-1}, \bftheta_t ]$ is given in \eqref{eq:filtlikelihood}.


\section{Application: Total Precipitable Water Measured by Three Sensor Systems \label{sec:application}}

We applied our methodology to hourly measurements from three sensor systems to obtain spatio-temporal filtering inference on an atmospheric variable called total precipitable water. Total precipitable water is the integrated amount of water vapor in a column from the surface of the earth to space in kilograms per square meter or, equivalently, in millimeters of condensate. The sensor systems are ground-based GPS, the Geostationary Operational Environmental Satellite (GOES) infrared sounders, and Microwave Integrated Retrieval System (MIRS) satellites. These data products are retrieved and stored at different data centers, and so our Situation 1 described in Section \ref{sec:intro} applies. The sensor systems also feature varying spatial coverage and precision. The measurement-error standard deviations are \mbox{$0.75$ mm}, \mbox{$2$ mm}, and \mbox{$4.5$ mm}, respectively, and so the function $v_{\epsilon}(\cdot)$ from \eqref{eq:datamodel} varies by server (i.e., by $j$) but not over space. 

Since March 2009, an operational blended multisensor water vapor product based on these three sensor systems has been produced by the National Environmental Satellite, Data, and Information Service of NOAA \citep{Kidder,Forsythe2012}. This product is sent to National Weather Service offices, where it is used by forecasters to track the movement of water vapor in the atmosphere and to detect antecedent conditions for heavy precipitation. The operational product is created by overlaying the existing field with the latest available data, which can lead to unphysical features in the form of abrupt boundaries. The goal of our analysis was to illustrate our methodology using a simple version of a spatio-temporal low-rank model, and to create spatially more coherent predictive maps with associated uncertainties based on data from all three systems, without having to transfer the data to a central processor.

\begin{figure}[H]
\large 
\hspace{18mm} $\mathbf{t=7}$ \hspace{4.2cm}  $\mathbf{t=8}$ \hspace{4.3cm} $\mathbf{t=9}$

\centering
\vspace{3mm}

\includegraphics[width=.28\textwidth]{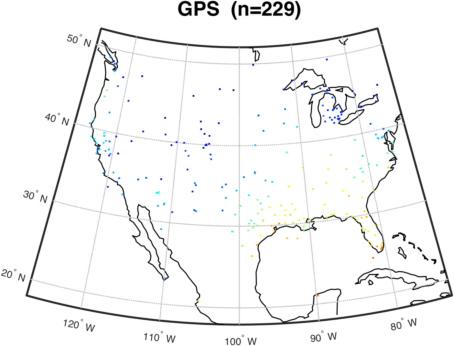} \hfill 
\includegraphics[width=.28\textwidth]{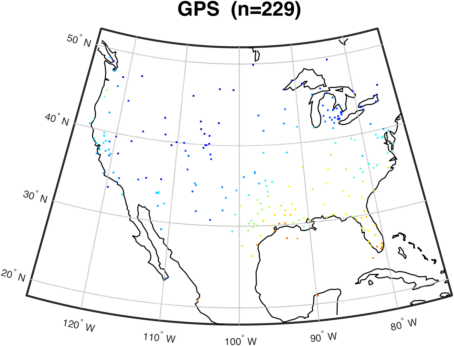} \hfill
\includegraphics[width=.28\textwidth]{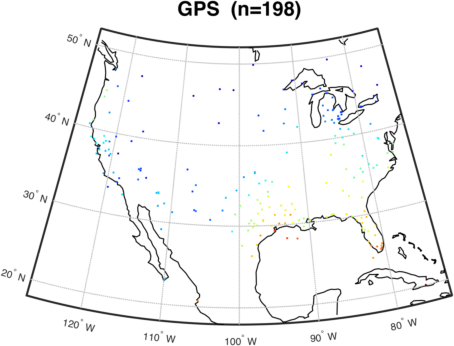}

\vspace{3mm}

\includegraphics[width=.28\textwidth]{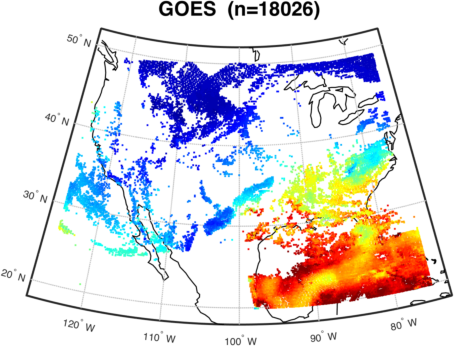} \hfill 
\includegraphics[width=.28\textwidth]{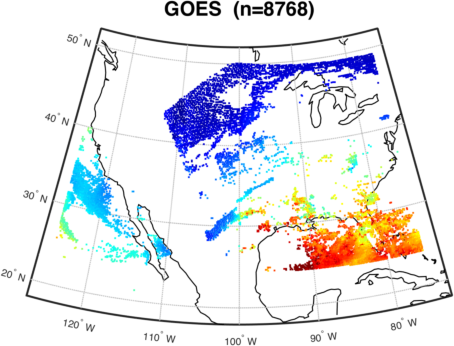} \hfill
\includegraphics[width=.28\textwidth]{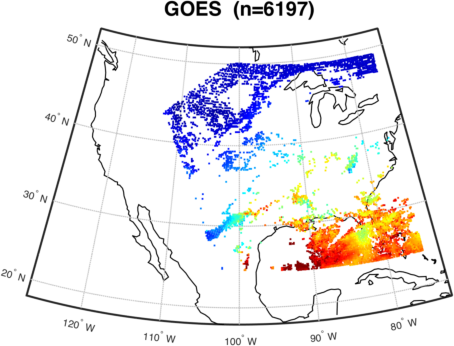}

\vspace{3mm}

\includegraphics[width=.28\textwidth]{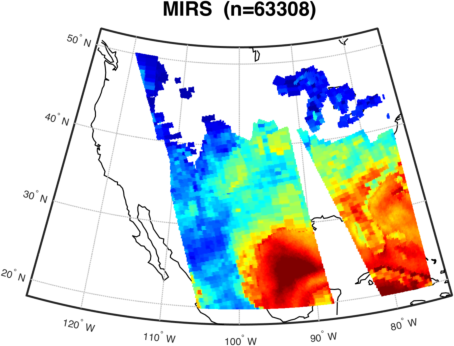} \hfill 
\includegraphics[width=.28\textwidth]{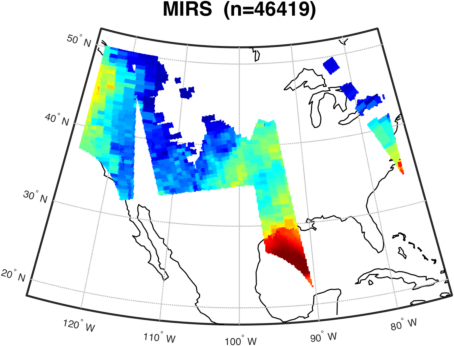} \hfill
\includegraphics[width=.28\textwidth]{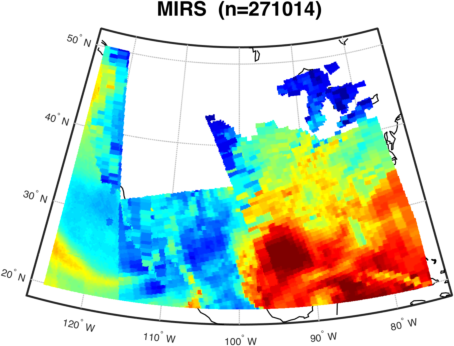}

\vspace{3mm}

\includegraphics[width=.28\textwidth]{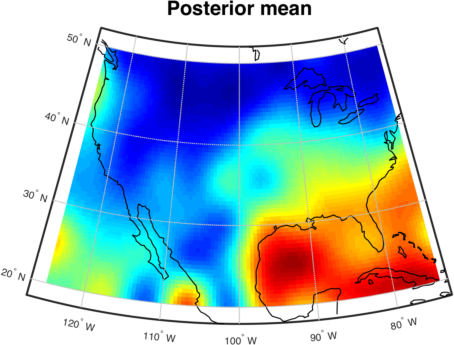} \hfill 
\includegraphics[width=.28\textwidth]{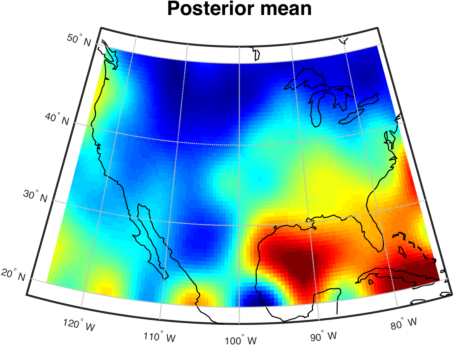} \hfill
\includegraphics[width=.28\textwidth]{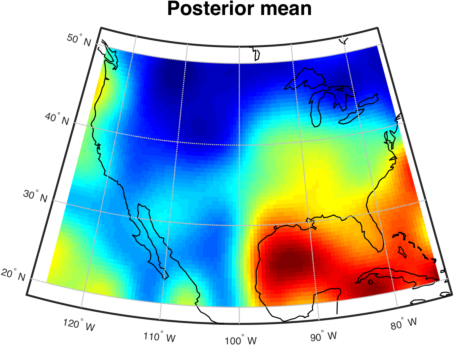}

\includegraphics[trim=0cm 6mm 0cm 10cm, clip=true,width=.7\textwidth]{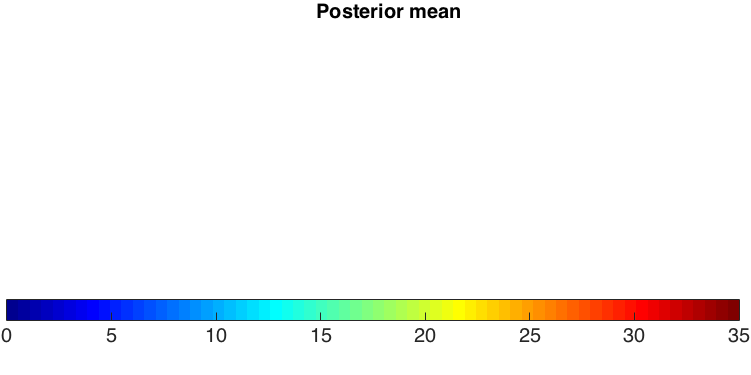}

\vspace{3mm}

\includegraphics[width=.28\textwidth]{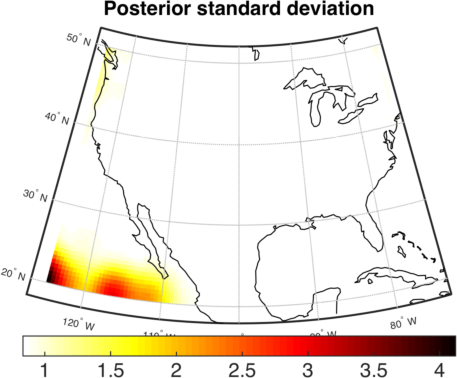} \hfill 
\includegraphics[width=.28\textwidth]{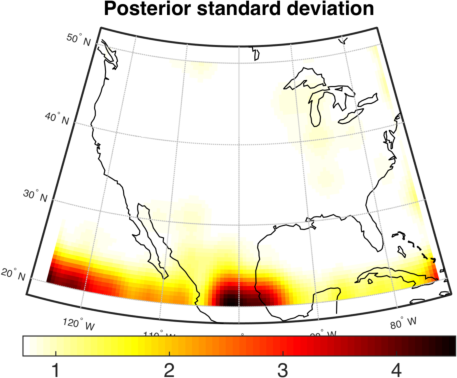} \hfill
\includegraphics[width=.28\textwidth]{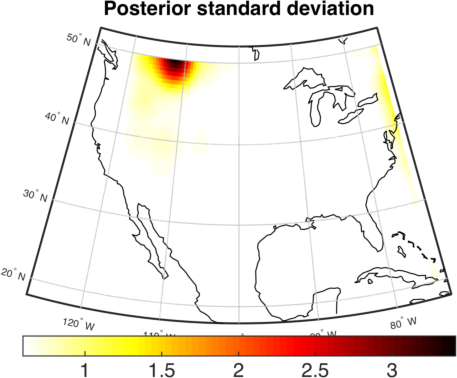}

%
%
%
%
%

\caption{Top three rows: Hourly observations of total precipitable water by the GPS system, GOES infrared sounders, and MIRS, respectively, over the larger continental United States in January 2011. Bottom two rows: Corresponding filtering posterior means and posterior standard deviations, respectively, of total precipitable water based on all three data products. The columns represent time points 7, 8, and 9, respectively. The scale for the posterior standard deviation plots varies between time periods. All units are in millimeters. 
}
\label{fig:results_animated_inline}
\end{figure}

We consider here a dataset consisting of a total of 3,351,860 measurements assumed to be collected at point-level support in January 2011 over a period of 47 hours by the three sensor systems over a spatial domain covering the United States. The top three rows of Figure \ref{fig:results_animated_inline} show the three sensor data products at time points (hours) 7, 8 and 9. As is evident from these plots, total precipitable water exhibits considerable variability at the considered spatial and temporal scales.

We made filtering inference based on a spatio-temporal low-rank model, parameterized by a predictive-process approach inspired by \citet{Finley2012}. Specifically, we assumed the model in Section \ref{sec:spatiotemporal} with $v_{\delta,t}(\cdot) \equiv \sigma^2_{\delta,t}$, $\bH_t = \alpha_t \bI_r$,
\begin{equation*}
\begin{split}
\bK_{0,0}^{-1} & = ( \rho( \bw_i, \bw_j | \bftheta_0))_{i,j = 1,\ldots,r}\\
\bU_t^{-1} & = (1 - \alpha_t^2)^{-1} ( \rho( \bw_i, \bw_j | \bftheta_t))_{i,j = 1,\ldots,r}\\
\bb_t(\bs) & = \sigma_t(\bs) \,\big(\rho( \bs, \bw_1 | \bftheta_t),\ldots, \rho( \bs, \bw_r | \bftheta_t)\big)', \; \bs \in \domain.
\end{split}
\end{equation*}
The parent correlation function was chosen to be
\[
\rho(\bs_1,\bs_2 | \bftheta_t ) = \mathcal{M}(\| \bs_1 - \bs_2 \|/\kappa_t)\cdot \mathcal{T}(\| \bs_1 - \bs_2 \|/10), 
\]
where $\mathcal{M}$ is the Mat\'ern correlation function \citep[e.g.,][p.\ 50]{Stein1999} with smoothness $\upsilon = 1.25$,
\[
\mathcal{M}(h) = ( 2 h \sqrt{\upsilon} )^\upsilon \mathcal{K}_\upsilon( 2 h \sqrt{\upsilon} ) 2^{1-\upsilon} / \Gamma(\upsilon)
\]
and multiplication by the compactly supported Kanter's function $\mathcal{T}$ \citep{Kanter1997} led to considerable sparsity in the matrices $\bR_{j,t}$, as described in Section \ref{sec:sparsity}. The set of knots, $\knots \colonequals \{ \bw_1, \ldots, \bw_{84} \}$, was a regular $5\dg \times 5\dg$ latitude/longitude grid over the domain. The trend consisted of an intercept term with a Gaussian random-walk prior with initial value 13.2 and variance 15.9 and was absorbed into the basis-function vector. While we chose this relatively simple model here for illustration, we would like to reiterate that neither the communication cost nor the computational complexity of the algorithm changes if a more elaborate parameterization of the general spatio-temporal low-rank model in Section \ref{sec:spatiotemporal} is chosen.

The transition distribution of the parameter vector 
\[
  \bftheta_t = (\Phi^{-1} (\alpha_t), \log(\sigma_t), \log(\kappa_t),\log(\sigma^2_{\delta,t}))'
\]
was taken to be a Gaussian random walk with $\bftheta_t|\bftheta_{t-1} \sim N_4(\bftheta_{t-1},0.01 \times \bI_4)$, for $t=1,\ldots,T=47$. The initial parameter vector $\bftheta_0$ was specified as $\alpha_0=0.8$, $\sigma_0=5$, $\kappa_0=15$, and $\sigma^2_{\delta,0}=0.5$. Here, $\alpha_t$ determines the strength of the temporal dependence, while the scale parameter $\kappa_t$ determines the strength of the spatial dependence. 

We implemented the sequential importance sampling algorithm with residual resampling as described in Algorithm \ref{alg:particlefilter} with $M=6000$ particles, using the prior distribution as the proposal distribution for simplicity. 
The resulting filtering posterior means and posterior standard deviations for total precipitable water for time periods 7, 8, and 9 (i.e., $t \in \{7,8,9\}$) on a regular $0.5\dg \times 0.5\dg$ latitude/longitude grid of size 6,283 are shown in the bottom two rows of Figure \ref{fig:results_animated_inline}. We were able to calculate the filtering distribution based on the 3,351,860 measurements collected over 47 hours by the three sensor systems in about 7 hours using parallel computations on the Geyser data analysis cluster on the high-performance computing facility Yellowstone at the National Center for Atmospheric Research. Geyser uses 10-core 2.4-GHz Intel Xeon E7-4870 (Westmere EX) processors and has 25GB of memory/core.

 
\section{Timing Study \label{sec:timing}} 

\begin{figure*}
\centering
\includegraphics[trim=6cm 9.5cm 6cm 10cm, clip=true,width=.48\textwidth]{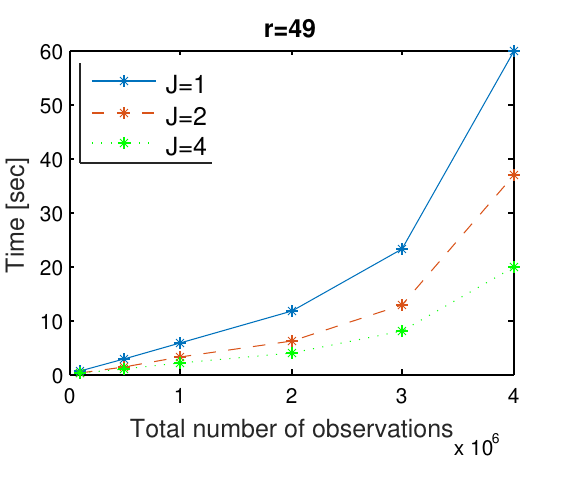} \hfill 
\includegraphics[trim=6cm 9.5cm 6cm 10cm, clip=true,width=.48\textwidth]{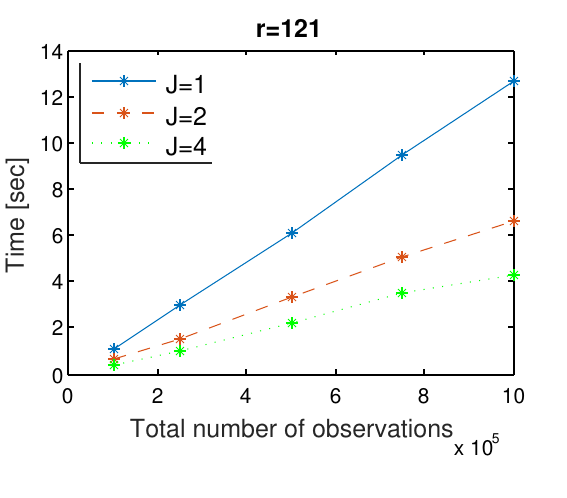} 
\caption{Computation time for one likelihood evaluation for simulated spatial data with a varying number of observations and different numbers of computational nodes ($J$), for $r=49$ (left) and $r=121$ (right) knots}
\label{fig:timing}
\end{figure*}

While the focus of this article is on avoiding data movement and duplicate storage for distributed data (Situation 1 from Section \ref{sec:intro}), the outlined methodology is also applicable without modification to a divide-and-conquer inference scheme in the case of centrally stored data (Situation 2). We briefly investigated the benefits in terms of computational speed by parallelizing one spatial-only likelihood evaluation for a predictive-process model with a Mat\'{e}rn covariance function similar to the one in Section \ref{sec:application}, for $r=49$ and $r=121$ knots, and varying numbers of simulated observations and numbers of servers.
The timing results shown in Figure \ref{fig:timing} are the means of ten replicates (the variation between replicates is very small). The study was conducted on a MacBook Pro with an Intel quad-core 2.6 GHz i7 processor and 8 GB of memory. The specific results are dependent on the characteristics of the processor, but provide a relative sense of computational improvement potential. We see that parallelizing over several processors leads to speed-ups, as expected. We would like to emphasize that, while not investigated further here, the divide-and-conquer scheme made possible by our methodology can also lead to memory advantages by splitting up the analysis in a distributed-memory environment. This is crucial when, for example, analyzing sea-surface temperature with hundreds of millions of measurements per day.

\section{Conclusions and Future Work \label{sec:conclusions}}

As datasets are becoming larger, so is the cost of moving them to a central computer for analysis, necessitating algorithms designed to work on distributed data that keep analysis operations as close to the stored data as possible. We showed how distributed spatial inference, including likelihood-based parameter inference, can be carried out in a computationally feasible way for massive distributed datasets under the assumption of a low-rank model, while producing the same results as traditional, non-distributed inference. Our approach is scalable in that the computational cost is linear in each $n_j$ (the number of measurements at server $j$) and the communication cost does not depend on the $n_j$ at all. Inference, especially when done based on a particle sampler, is also ``embarassingly parallel,'' allowing a divide-and-conquer analysis of massive spatial data with little communication overhead. In addition, if the selected low-rank model has fixed basis functions that do not depend on parameters (see Section \ref{sec:sreinference}), our methodology can be used for data reduction in situations where it is not possible to store all measurements.

After extending the results to the spatio-temporal case, we demonstrated the applicability of our model to massive real-world data in Section \ref{sec:application}, and showed that we can obtain sensible results in a fast manner. However, getting the best possible results for this particular application is part of ongoing research and will likely require a more refined and complicated model.

The methodology described in this article can be extended to the full-scale approximation of \citet{Sang2011a}, where the fine-scale variation is assumed to be dependent within subregions of the spatial domain, resulting in nondiagonal $\bV_j$.
This idea is explored for a multi-resolutional extension of the full-scale approximation in \citet{Katzfuss2015}.

Another natural extension of our methodology is to the increasingly important multivariate data-fusion case involving inference on multiple processes based on data from multiple measuring instruments. Multivariate analysis can in principle be carried out as described here by stacking the basis function weights for the individual processes into one big vector $\bfeta$ \citep[see, e.g.,][]{Nguyen2011,Nguyen2012}, but it will likely require more complicated inference on $\delta(\cdot)$ due to different instrument footprints and overlaps. While the combined size of the low-rank components for multiple processes will become prohibitive in highly multivariate settings, the hope is that the processes can be written as linear combinations of a smaller number of processes.

\section*{Acknowledgements}
This material was based upon work partially supported by the National Science Foundation under Grant DMS-1127914 to the Statistical and Applied Mathematical Sciences Institute. Katzfuss was partially supported by NASA's Earth Science Technology Office AIST-14 program and by National Science Foundation (NSF) Grant DMS-1521676. Hammerling's research also had partial support from the NSF Research Network on Statistics in the Atmosphere and Ocean Sciences (STATMOS) through grant DMS-1106862. We would like to acknowledge high-performance computing support from Yellowstone (ark:/85065/d7wd3xhc) provided by NCAR's Computational and Information Systems Laboratory, sponsored by the National Science Foundation. We would like to thank Amy Braverman for making us aware of the problem of distributed spatial data; John Forsythe and Stan Kidder for the datasets and helpful advice; Yoichi Shiga for support with preprocessing and visualizing the data; and Andrew Zammit Mangion, Emtiyaz Khan, Kirk Borne, Jessica Matthews, Emily Kang, several anonymous reviewers, and the SAMSI Massive Datasets Environment and Climate working group for helpful comments and discussions.

\appendix

\section{Derivation of the Likelihood \label{appendix}}

We derive here the expression of the likelihood in \eqref{eq:distributedlikelihood}. First, note that $\bz_{1:J} | \bftheta \sim N_n(\bB_{1:J}\bfnu_0,\bfSigma_{1:J})$, where $\bfSigma_{1:J} = \bB_{1:J} \bK_0 \bB_{1:J}' + \bV_{1:J}$. Hence, the likelihood is given by,
\[
 -2 \log [\bz_{1:J} | \bftheta ] = \log|\bfSigma_{1:J}| + (\bz_{1:J} - \bB_{1:J}\bfnu_0)'\bfSigma_{1:J}^{-1}(\bz_{1:J} - \bB_{1:J}\bfnu_0) -(n/2)\log(2\pi).
\]
Applying a matrix determinant lemma \citep[e.g.,][Thm.~18.1.1]{Harville1997}, we can write the log determinant as,
\begin{align*}
  \log|\bfSigma_{1:J}| & = \log |\bV_{1:J}| + \log|\bK_0| + \log| \bB_{1:J}'\bV_{1:J}^{-1} \bB_{1:J} + \bK_0^{-1} | \\
      & = \textstyle \sum_{j=1}^J \log|\bV_j| - \log|\bK_0^{-1}| + \log|\bK_z^{-1}|.
\end{align*}
Further, using the Sherman-Morrison-Woodbury formula \citep{Sherman1950,Woodbury1950,Henderson1981}, we can show that $\bfSigma_{1:J}^{-1} = \bV_{1:J}^{-1} - \bV_{1:J}^{-1} \bB_{1:J} \bK_z \bB_{1:J}' \bV_{1:J}^{-1}$, and so
\begin{align*}
 &  (\bz_{1:J} - \bB_{1:J}\bfnu_0)'\bfSigma_{1:J}^{-1}(\bz_{1:J} - \bB_{1:J}\bfnu_0)  \\
 & = \textstyle \sum_{j=1}^J (\bz_j - \bB_j\bfnu_0)'\bV_j^{-1}(\bz_j - \bB_j\bfnu_0) \\
    & \quad \textstyle - \big(\sum_{j=1}^J \bB_j'\bV_j^{-1}(\bz_j-\bB_j\bfnu_0)\big)'\bK_z \big(\sum_{j=1}^J \bB_j'\bV_j^{-1}(\bz_j-\bB_j\bfnu_0)\big)\\
 & = \textstyle \sum_j \bz_j'\bV_j^{-1} \bz_j -2 \bfnu_0'(\bK_z^{-1}\bfnu_z-\bK_0^{-1}\bfnu_0)+ \bfnu_0'(\bK_z^{-1}- \bK_0^{-1}) \bfnu_0 \\
    & \quad \textstyle  - \big( (\bK_z^{-1}\bfnu_z-\bK_0^{-1}\bfnu_0)-(\bK_z^{-1}-\bK_0^{-1})\bfnu_0\big)'\bK_z \big( (\bK_z^{-1}\bfnu_z-\bK_0^{-1}\bfnu_0)-(\bK_z^{-1}-\bK_0^{-1})\bfnu_0\big)\\
& = \textstyle \sum_j \bz_j'\bV_j^{-1} \bz_j - 2\bfnu_0'\bK_z^{-1}\bfnu_z + \bfnu_0'\bK_0^{-1}\bfnu_0 + \bfnu_0'\bK_z^{-1}\bfnu_0 \\
    & \quad \textstyle - (\bK_z^{-1}\bfnu_z)'\bK_z(\bK_z^{-1}\bfnu_z)-\bfnu_0'\bK_z^{-1}\bK_z\bK_z^{-1}\bfnu_0 + 2(\bK_z^{-1}\bfnu_z)'\bK_z \bK_z^{-1} \bfnu_0\\
& = \textstyle  \sum_j \bz_j'\bV_j^{-1} \bz_j + \bfnu_0'\bK_0^{-1}\bfnu_0 - \bfnu_z'\bK_z^{-1}\bfnu_z,
\end{align*}
where $\sum_{j=1}^J \bB_j'\bV_j^{-1}\bB_j = \bK_z^{-1}-\bK_0^{-1}$ and $\sum_{j=1}^J \bB_j'\bV_j^{-1}\bz_j = \bK_z^{-1}\bfnu_z-\bK_0^{-1}\bfnu_0$ both follow from \eqref{eq:infofilter}.

\section{Spatial Prediction When Observed and Prediction Locations Coincide \label{sec:predoverlap}}

Here we describe how to do spatial prediction when a small number, $q$ say, of the observed locations are also in the set of desired prediction locations. Define $\bfdelta_{P,O}$ to be the vector of the first $q$ elements of $\bfdelta^P$, which we assume to correspond to the $q$ observed prediction locations, and let $\bP_j$ be a sparse $n_j \times q$ matrix with $(\bP_j)_{k,l} = I(\bs_{j,k} = \bs_l^P)$. 
We write our model in state-space form with identity evolution equation,
$\bz_j = \tilde{\bB}_j \tilde{\bfeta} + \tilde{\bfxi}_j$,
where
$\tilde{\bB}_j \colonequals (\bB_j, \bP_j)$, 
$ \tilde{\bfeta} \colonequals (\bfeta', \bfdelta_{P,O}')' \sim N(\tilde{\bfnu}_0,\tilde{\bK}_0)$,  $\tilde{\bfnu}_0 \colonequals (\bfnu_0', \bfzero_q')'$, $\tilde{\bK}_0 $ is blockdiagonal with first block $\bK_0$ and second block $\diag\{ v_\delta(\bs^P_1),\ldots,v_\delta(\bs^P_q)\}$,
$\tilde{\bfxi}_j \sim N_{n_j}(\bfzero,\tilde{\bV}_j)$, and $\tilde{\bV}_j$ is the same as $\bV_j$ except that the $i$th diagonal element is now $v_\epsilon(\bs_{j,i})$ if $\bs_{j,i}$ is one of the prediction locations. 

The decentralized Kalman filter \citep{Rao1993} gives
$\tilde{\bK}_z^{-1} = \tilde{\bK}_{0}^{-1} + \textstyle\sum_{j=1}^J \tilde{\bR}_j$ and 
$\tilde{\bfnu}_z = \tilde{\bK}_z(\tilde{\bK}_0^{-1} \tilde{\bfnu}_0 + \textstyle\sum_{j=1}^J \tilde{\bfgamma}_j)$,
where $\tilde{\bR}_j \colonequals \tilde{\bB}_j'\tilde{\bV}_j^{-1} \tilde{\bB}_j$ and $\tilde{\bfgamma}_j \colonequals \tilde{\bB}_j'\tilde{\bV}_j^{-1}\bz_j$ are the only quantities that need to be calculated at and transfered from server $j$, which is feasible due to sparsity if $q$ is not too large. 
The predictive distribution is then given by $\by^P | \bz_{1:J} \sim N( \tilde{\bB}^P \tilde{\bfnu}_J, \tilde{\bB}^P \tilde{\bK}_z \tilde{\bB}^P{}' + \tilde{\bV}_\delta^P )$, where $\tilde{\bB}^P \colonequals (\bB^P, (\bI_q, \bfzero)')$ and $\tilde{\bV}_\delta^P \colonequals \diag\{\bfzero_q', v_\delta(\bs^P_{q+1}),\ldots,v_\delta(\bs^P_{n_P})\}$.

\bibliographystyle{apalike}
\bibliography{library}

\end{document}